\documentclass[journal=jctcce,manuscript=letter,preprint,amsmath,amssymb,bibtex]{achemso}

\usepackage{chemformula} 
\usepackage[T1]{fontenc} 

\usepackage{graphicx}
\usepackage{dcolumn}
\usepackage{bm}
\usepackage{setspace}
\usepackage{ulem}
\usepackage{color}
\usepackage{braket}
\usepackage{tabularx}
\usepackage{booktabs}
\usepackage{multirow}
\usepackage{caption}
\usepackage[subrefformat=parens]{subcaption}

\author{Mauro Cainelli}
\affiliation{Department of Chemistry, Graduate School of Science, Kyoto University, Kitashirakawa Oiwake-cho, Sakyo-ku Kyoto, 606-8502, Japan}
\email{cainelli@theoc.kuchem.kyoto-u.ac.jp}

\author{Reo Baba}
\affiliation{Department of Chemistry, Graduate School of Science, Kyoto University, Kitashirakawa Oiwake-cho, Sakyo-ku Kyoto, 606-8502, Japan}

\author{Yuki Kurashige}
\affiliation{Department of Chemistry, Graduate School of Science, Kyoto University, Kitashirakawa Oiwake-cho, Sakyo-ku Kyoto, 606-8502, Japan}
\alsoaffiliation{FOREST, JST, Honcho 4-1-8, Kawaguchi, Saitama 332-0012, Japan}
\altaffiliation{CREST, JST, Honcho 4-1-8, Kawaguchi, Saitama 332-0012, Japan}
\email{kura@kuchem.kyoto-u.ac.jp}

\title{Numerical investigation of the quantum inverse algorithm on small molecules}

\begin{document}

\maketitle

\begin{abstract}
We evaluate the accuracy of the quantum inverse (Q-Inv) algorithm in which the multiplication of $\hat{H}^{-k}$ to the reference wavefunction is replaced by the Fourier Transformed multiplication of $e^{-i\lambda \hat{H}}$, as a function of the integration parameters ($\lambda$) and the power $k$ for various systems, including H$_2$, LiH, BeH$_2$ and the notorious H$_4$ molecule at single point. We further consider the possibility of employing the Gaussian-quadrature rule as an alternate integration method and compared it to the results employing trapezoidal integration. The Q-Inv algorithm is compared to the inverse iteration method using the $\hat{H}^{-1}$ inverse (I-Iter) and the exact inverse by lower-upper decomposition (LU). Energy values are evaluated as the expectation values of the Hamiltonian. Results suggest that the Q-Inv method provides lower energy results than the I-Iter method up to a certain $k$, after which the energy increases due to errors in the numerical integration that are dependent of the integration interval. A combined Gaussian-quadrature and trapezoidal integration method proved to be more effective at reaching convergence while decreasing the number of operations. For systems like H$_4$, in which the Q-Inv can not reach the expected error threshold, we propose a combination of Q-Inv and I-Iter methods to further decrease the error with $k$ at lower computational cost. Finally, we summarize the recommended procedure when treating unknown systems.
\end{abstract}

\section{Introduction}

Quantum computing has been of great interest in the past years because it allows the direct treatment of state superposition and entanglement in the Hilbert space due to the employment of qubits, which could lead to exponential increase in calculation time when compared to classical computers.\cite{bauer2020chan, nielsen2010chuang, lloyd1996lloyd, mcardle2020yuan}
In practice, however, despite the advance in the design of quantum chips that might be used to study systems of the range of several tens qubits, the current algorithms are much more limited due to the increase in the number of operations, the number of quantum gates and the overall complexity of the circuit design with increasing qubit size.\cite{cao2019aspuru-guzik, wecker2014troyer}
Recent interests include the design of short depth algorithms which can solve problems in the noisy intermediate scale quantum (NISQ) devices.\cite{preskill2018preskill, sachdeva2013vishnoi, matsuzawa2020kurashige, berry2014somma, berry2015somma, childs2012wiebe, babbush2018chan, daskin2012kais, low2019wiebe}. These devices with large number of qubits are sensitive to the environment, whose interactions produce quantum decoherence, introducing noise into calculations which often requires the use of error corrections and mitigation techniques.\cite{kitaev1997kitaev, yoshioka2022endo}

In the present, there are fundamentally two main algorithms that can be used, and recently combined, to compute the ground state properties of the system Hamiltonian. \cite{cohn2021parrish, omalley2016martinis, klymko2022tubman}
The first one is the phase estimation algorithm (PEA), which consists in the use of unitary operators to compute the quantum dynamics of the system that is controlled by register qubits\cite{abrams1999lloyd, aspuru-guzik2005head-gordon,dorner2009walmsley, whitfield2011aspuru-guzik}, where the eigenvalue is encoded on relative phases of ancilla qubits.\cite{daskin2017kais} This algorithm usually give logarithmic error and polynomial gate scaling, but requires substantial circuit depth or number of ancillas for currently available circuits.
The second one is the variational quantum eigensolvers (VQE)\cite{bierman2023lu, kandala2017gambetta}, which is a hybrid quantum-classical method that employs quantum computers to calculate the energy expectation values of prepared trial quantum states (ansatz) that have the parameters variationally optimized by classical computers.\cite{lee2021babbush, park2024killoran, peruzzo2014obrien, huggins2020whaley, mcclean2017dejong} This method reduces the circuit depth requirement compared to the PEA, but increases the number of required measurements of trial states in order to achieve the same desired precision.\cite{wecker2015troyer, nishio2023kurashige} The Schrödinger equation is then solved only in the space defined by the ansatz.

Regarding the first routine, according to Ding et al.\cite{ding2023lin}, the performance of the quantum phase estimation (QPE) algorithms can be evaluated in terms of the number of ancilla qubits required, the maximum runtime $T_{\text{max}}$, measured by the maximum depth of the circuit, the total runtime $T_{\text{total}}$, equal to the sum of the circuit depths over all executions, and the minimal overlap $p_0 = |c_{0}|^2$ required between the initial state $\ket{\psi}$ and the target state $\ket{\phi_0}$. They suggested that a near-optimal phase estimation algorithm should meet the following requirements: 
(1) Use a small number of (even a single) ancilla qubits. 
(2) Allow the initial state $\ket{\psi}$ to be inexact. 
(3) Achieve the Heisenberg-limited scaling $T_{\text{total}} = O(\epsilon^{-1})$. 
(4) $T_{\text{max}} = O(\epsilon^{-1})$, and the prefactor can be arbitrarily small when $\ket{\psi}$ approaches to the exact eigenstate $\ket{\psi_0}$.

Achieving a small $T_{\text{max}}$ is particularly important in the design of current algorithms that treat systems with a small number of qubits and a relatively short coherence time.\cite{ni2023ying, lin2022tong}.
Quantum versions of the Lanczos method\cite{lanczos1950lanczos, kirby2023mezzacapo}, which is based on the application of powers of the Hamiltonian\cite{seki2021yunoki} to some reference state represented in a Krylov space, have been studied to approximate real\cite{cohn2021parrish, cortes2022gray, stair2020evangelista, parrish2019mcmahon, klymko2022tubman} an imaginary\cite{kamakari2022minnich, motta2020chan} time evolutions.
In the same direction, the quantum inverse (Q-Inv) algorithm for the estimation of the ground state energy (GSE) of a quantum system was first proposed by Kyriienko\cite{kyriienko2020kyriienko}. This method applies the power of the inverse Hamiltonian to the state and it is inspired by the classical inverse iteration (I-Iter) algorithm for finding the dominant eigenstate of the matrix. It falls in the category of quantum Fourier transform (QFT) methods,\cite{kitaev1995kitaev} where it extends the Fourier approximation proposed by Childs et al.\cite{childs2017somma} to the $k^{\text{th}}$ power of the inverse, providing exponential speed up for the inverse Hamiltonian multiplication process that is performed by a quantum circuit. A modified version of the Q-Inv method was also recently proposed by Yoshikura et al. for the determination of general eigenstates.\cite{yoshikura2023tsuchimochi}    
In Kyriienko's work, results of the H$_2$ and BeH$_2$ systems showed that, at least for those particular small systems, the approximation of the energy works well up to a certain $k^{\text{th}}$ order, after which the energy increases and deviates from the ideal inverse, while staying below the error threshold. Although it seemed that the limitation of $k$ depended on the defined maximum phase ($\phi_{\text{max}}$), it was not exactly clear why this behaviour occurred. Furthermore, calculations were conducted employing the trapezoidal rule for different integration ranges and interval sizes reflected in the global $\phi_{\text{max}}$. While there were some calculations that showed the relation between the discretization intervals through the called "skew" parameter, there still seems to be a lack of information regarding the optimization of the integration parameters for different systems and comparison with different integration methods.

In this work, we assess the parameters that affect the accuracy of the Q-Inv method on various systems, including the notorious H$_4$ molecule at the singular point.\cite{lee2019whaley} In some cases where the method cannot reach the error threshold, we focus on the idea of using the Q-Inv algorithm to compute the inverse up to a maximum $k^{\text{th}}$ order with the minimum error, and then employ the iterative (I-Iter) procedure to further decrease of the error to the arbitrary precision. In the systems we study, we show that after the $k^{\text{th}}$ order estimation, only a couple of iteration steps might be enough to estimate the energy under the precision threshold. We evaluate the possibility of employing different integration rules\cite{gauß1814methodus, murray1993laming} to further decrease the number of operator multiplications employed in the algorithm.

\section{Methods}

The inverse iteration method is based on the premise that the quantum linear system problem (QLSP) $\hat{A}\overrightarrow{x}=\overrightarrow{b}$ is equivalent to applying the inverse operator $\hat{A}^{-1}$ to the state $\ket{b}$.\cite{harrow2009lloyd} Because $\hat{A}^{-1}$ is a nonunitary operator, the strategy is to represent it as a linear combination of unitaries. The general choice is exponentials of the type $e^{-i\hat{A}t}$ that can be implemented by any algorithm.\cite{aaronson2015aaronson} Thus, $\hat{A}^{-1}$ can be written as $\hat{A}^{-1}=\int_{0}^{\infty}e^{-i\hat{A}t}dt$.
The Fourier approach proposed by Childs et al.\cite{childs2017somma} expresses the exponential integral as a double integral of the propagator $t$ and approximate it by finite sums. For the problem of finding the eigenenergies of a system, the operator $\hat{A}$ corresponds to the Hamiltonian and the Fourier approximation for the decomposition of $t$ into $y$ and $z$ variables is expressed:

\begin{equation}
\hat{H}^{-1}=\frac{i}{\sqrt{2\pi}}\int_{0}^{\infty}dy\int_{-\infty}^{\infty}dz ze^{-\frac{z^2}{2}}e^{-iyz\hat{H}}
\label{org_inv}
\end{equation}

The approximation was recently extended by Kyriienko to the $k^{\text{th}}$ power of the inverse, for $k\ge$1.\cite{kyriienko2020kyriienko}

\begin{equation}
\hat{H}^{-k}=\frac{iN_k}{\sqrt{2\pi}}\int_{0}^{\infty}dy\int_{-\infty}^{\infty}dz zy^{k-1}e^{-\frac{z^2}{2}}e^{-iyz\hat{H}}
\label{gen_inv}
\end{equation}
where $N_k$ is a normalization factor. For $k=1$ Eq. \ref{gen_inv} reduces to Eq. \ref{org_inv}. The discretization of the integral leads to 
\begin{equation}
\begin{split}
\hat{H}^{-k} \approx &\frac{iN_k}{\sqrt{2\pi}}\sum_{j_y=0}^{M_y-1}\Delta_y\sum_{j_z=-M_z}^{j_z=M_z}\Delta_z (j_z\Delta_z)\\
             &       (j_y\Delta_y)^{k-1}e^{-\frac{{j_z}^2{\Delta_z}^2}{2}}e^{-i(j_z\Delta_z)(j_y\Delta_y)\hat{H}}
\label{disc_inv}
\end{split}
\end{equation}
where $\Delta_{y}$ and $\Delta_{z}$ correspond to the discretization steps, while $M_{y}$ and $M_{z}$ represent the cutoffs of the integration ranges for the $y$ and $z$ variables, respectively.
While the sum of the discretized variables can be done by employing the trapezoidal rule, here we explore the possibility of using the Gaussian-quadrature rule to reduce the number of iterations and the number of matrix multiplications required by the algorithm.\cite{murray1993laming}

The Gaussian-quadrature rule approximates the function $f(x)$ by a choice of nodes ($x_i$) and weights ($w_i$) usually in the domain $[-1,1]$, for $i=1,...,n$ and it gives the exact result for polynomials of degree $2n-1$ or less. This particular rule is the Gauss-Legendre rule and the weights are associated to the orthogonal Legendre polynomials.\cite{gauß1814methodus}

\begin{equation}
\int_{-1}^{1}f(x)dx \approx \sum_{i=1}^{n}w_if(x_i)
\end{equation}
Applying the Gauss-Legendre rule to the quantum inverse algorithms requires a change of the integration interval to the $[-1,1]$ domain, which can be done in the following way.

\begin{equation}
\int_{a}^{b}f(x)dx \approx \frac{b-a}{2}\sum_{i=1}^{n}w_if\bigg(\frac{b-a}{2}x_i + \frac{a+b}{2}\bigg)
\label{int_change}
\end{equation}
After using Eq. \ref{int_change} to change the integration interval of both $y$ and $z$ variables in Eq. \ref{gen_inv}, the discrete sum read as follows.
\begin{equation}
\begin{split}
\hat{H}^{-k} \approx &\frac{iN_k}{\sqrt{2\pi}}c_{a,b,c,d}\sum_{j_y=1}^{n_y}w_{j_y}\sum_{j_z=1}^{n_z}w_{j_z}{z'}_{j_z}{y'}_{j_y}^{k-1}\\ 
       &\exp\bigg(-\frac{{z'}_{j_z}^2}{2}\bigg)\exp\bigg(-iz'_{j_z}y'_{j_y}\hat{H}\bigg)
\label{gen_legendre}
\end{split}
\end{equation}
where $z'_{j_z}=dz_{j_z}$ and $y'_{j_y}=\Big(\frac{b}{2}y_{j_y}+\frac{b}{2}\Big)$ are the re-defined variables in terms of the Legendre polynomials $z_{j_z}$ and $y_{j_y}$ with the intervals $[0,b]$ and $[-d,d]$, respectively. The term $c_{a,b,c,d}=\frac{b}{2}d$ is the constant for the change of interval, $w_{j_z}$ and $w_{j_y}$ are the weights of the Legendre polynomials with order $n_z$ and $n_y$ used in the approximation of $z$ and $y$, respectively. Note that Eq. \ref{gen_legendre} can be used with other quadrature methods by changing the definition of the variables and intervals.
A more compact notation can be obtained as stated in Ref. \citenum{kyriienko2020kyriienko}.

\begin{equation}
\hat{H}^{-k} \approx \sum_{b}\sum_{d}c_{y,z,k}\exp\bigg(-iz'_{j_z}y'_{j_y}\hat{H}\bigg)
\label{com_legendre}
\end{equation}
where $c_{y,z,k}$ represent purely imaginary coefficients for the series. We point out that in Eq. \ref{com_legendre}, the exponential variables and the summations does not explicitly depend on k, which means that, in principle, we can compute any k$^{th}$ order Hamiltonian at the same computational cost.

In order to study some systems with large number of qubits, we use the complete active space (CAS) model,\cite{taylor1980siegbahn, takeshita2020mcclean} in which the molecular orbitals are divided into inactive orbitals, which are low energy orbitals occupied by electrons, virtual orbitals, which are high energy unoccupied orbitals, and the active orbitals, which are orbitals that can be occupied or unoccupied. The CAS calculation is performed on the active orbitals reduced space, while the inactive and virtual orbitals are considered frozen and included in a constant base energy (\text{$E_{0}$}). \cite{nishio2023kurashige}
\begin{equation}
\begin{split}
\hat{H}_\text{CAS} &= \hat{H}_\text{act}+E_{0}\\
&= \sum_{pq,\sigma}h_{pq,\sigma}{a^{\dag}_{p\sigma}}a_{q\sigma}
+ \frac{1}{2}\sum_{pqrs,\sigma\tau}g_{pqrs}a^{\dag}_{p\sigma}a^{\dag}_{r\tau}a_{s\tau}a_{q\sigma} + E_{0}
\end{split}
\end{equation}
where $h_{pq}$ and $g_{pqrs}$ are the one-electron and two-electron integrals, respectively, $a^{\dag}_{p\sigma}$ and $a_{p\sigma}$ are the creation and an annihilation operators for the spatial orbital $p$ and spin label $\sigma$. The active energy ($E_{\text{act}}$) using the Q-Inv method is obtained by approximating the $k^{\text{th}}$ inverse ($\hat{H}^{-k}_{\text{act}}$) in Eq. \ref{gen_legendre} and computing the expected value as

\begin{equation}
E_{\text{act}} = \frac{\bra{\hat{H}^{-k}_{\text{act}}\hat{\Psi}_{0}}\hat{H}_{\text{act}}\ket{\hat{H}^{-k}_{\text{act}}\hat{\Psi}_{0}}}{\braket{\hat{H}^{-k}_{\text{act}}\hat{\Psi}_{0}|\hat{H}^{-k}_{\text{act}}\hat{\Psi}_{0}}}
\end{equation}
where $\hat{\Psi}_{0}$ is the initial guess of the CAS wavefunction. Then, the CAS energy ($\varepsilon_{\text{CAS}}$) can be approximated by the addition of the constant energy to the active energy.

\begin{equation}
\varepsilon_{\text{CAS}} \le E_{\text{CAS}} = E_{\text{act}} + E_{0}.
\label{approx_ener}
\end{equation}

The $k^{\text{th}}$ order inverse Hamiltonian can also be obtained using the iteration method by approximating $\hat{H}_{\text{act}}^{-1}$ in Eq. \ref{gen_legendre} with $k$=1 and applying the inverse $k$-times to the wavefunction.

\begin{equation}
\hat{H}_{\text{act}}^{-k}\hat{\Psi}_{0} = \Big(\hat{H}_{\text{act}}^{-1}\Big)^{k}\hat{\Psi}_{0}
\label{H_act_inv}
\end{equation}

The advantage of the Q-Inv method proposed by Kiiryenko\cite{kyriienko2020kyriienko} (Eq. \ref{gen_inv}) over the I-Inv method by Childs et al.\cite{childs2017somma} (Eq. \ref{org_inv}) where the multiplication of $\hat{H}^{-k}$ is realized by the multiplication of $\hat{H}^{-1}$ $k$-times is, then, the possibility of reducing the number of operator multiplications of the order $O(k)$ by directly approximating $\hat{H}^{-k}$. However, as discussed in the original work, this approximation might present some limitations as the iteration power $k$ increases. Both methods of calculation will also be analyzed and compared here.

\section{Results}

We calculated the ground state energy for the H$_2$, LiH, BeH$_2$ and H$_4$ systems using the Q-Inv method. The initial guess of the ground state was prepared to be the Hartree-Fock (HF) state ($\hat{\Psi}_{0} = \hat{\Psi}_{\text{HF}}$) and the Hamiltonian in Eq. \ref{gen_legendre} was shifted by the HF energy ($\hat{H}_{\text{act}}-E_{\text{HF}}$) in order to make the initial state well conditioned by increasing the overlap with the ground state and reducing the overlap with the excited states.

The HF states were obtained using the openfermion pyscf packages.\cite{mcclean2019babbush, sun2018chan, sun2020chan} In the case of H$_2$ and H$_4$, we computed the Full-CI energies corresponding to 4 and 8 qubit systems, respectively. For LiH and BeH$_2$, we employ the CAS-CI method to reduce the number of qubits from 12 and 14, respectively, to 10 qubits each by excluding the 1s orbitals of Li and Be, and the $\sigma_u^*$ anti-bonding orbital in the BeH$_2$ case. The H--H bond distances for H$_2$ and H$_4$ were set to 0.75 {\AA} and 1.23 {\AA}, respectively. The Li--H bond distance in LiH was set to 1.6 {\AA} and the Be--H bond distance in BeH$_2$ was set to 1.326 {\AA}. All calculations were performed using the STO-6G basis set.\cite{10.1063/1.1672392}

\subsection{Discretization Parameters}

The integration parameters were determined in the following order:
(1) Determine the appropriate integration cutoff intervals $[0,b]$ and $[-d,d]$ for the variables $y$ and $z$, respectively.
(2) Determine the minimum order of Legendre polynomials/discrete intervals needed for the integration along $y$ ($n_y$) for large enough $n_z$.
(3) Determine the minimum order of the Legendre polynomials/discrete intervals needed for the integration along $z$ ($n_z$) for the minimum $n_y$ obtained.
(4) Determine the minimum iteration step $k$ needed to account for an error below the tolerance $\epsilon$.

\subsubsection{Integration Intervals}
\begin{figure*}[!t]
\center{\includegraphics[width=\columnwidth, keepaspectratio]
        {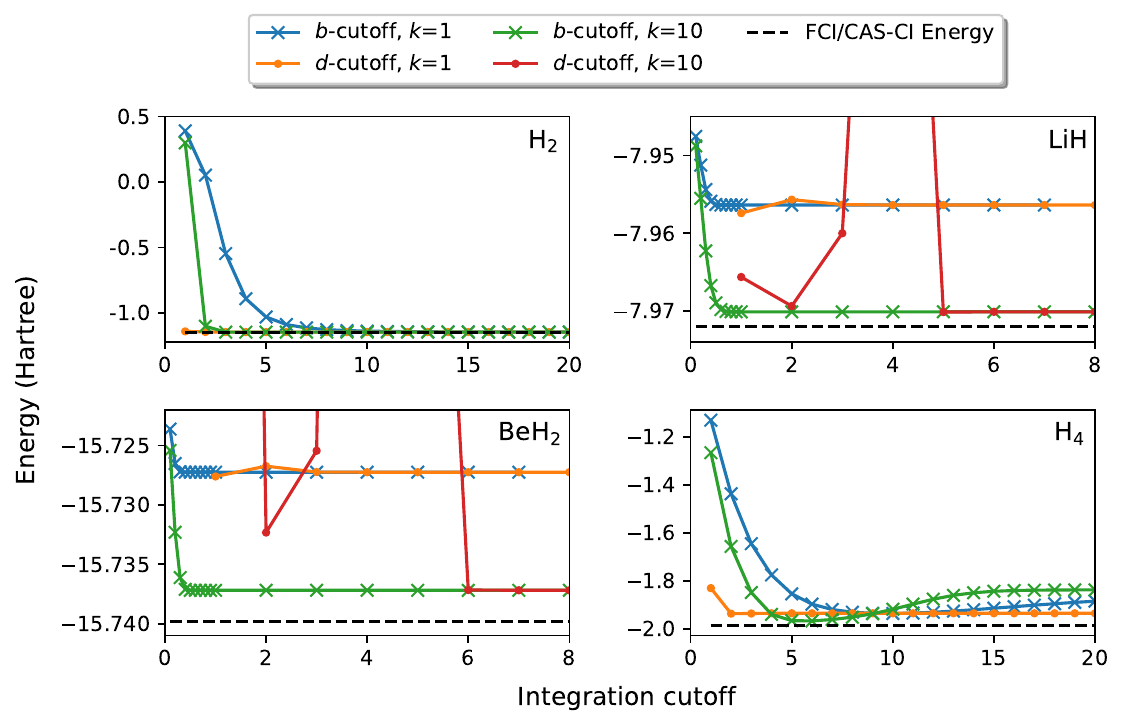}}
        \caption{Ground state energy as a function of the integration cutoff along $y$ ($b$-cutoff) and along $z$ ($d$-cutoff) at two different iteration steps ($k$=1 and $k$=10) for H$_2$ (top left), LiH (top right), BeH$_2$ (bottom left) and H$_4$ (bottom right). The FCI energy for H$_2$ and H$_4$, and the CAS-CI energy for LiH and BeH$_2$, are represented as black dashed lines.}
        \label{fig1}
\end{figure*}

Although the order of the integration of $y$ and $z$ can be reversed, here we compute the integration along $z$ first. Integrations are performed using the Gauss-Legendre quadrature rule.
The number of Legendre polynomials in Eq. \ref{gen_legendre} are initially set to $n_y$=50 and $n_z$=300 for all calculations, which are large enough to make any discretization errors negligible, and results are plotted for two iteration step values ($k$=1 and $k$=10). The interval $[-d,d]$ for $z$ is chosen to be $[-5,5]$ in the calculation of the cutoff for $b$ ($b$-cutoff). Then we use the $b$-cutoff value for the interval $[0,b]$ for $y$ to calculate the cutoff for $d$ ($d$-cutoff).
The obtained energies $E_{\text{CAS}}$ are shown in Fig. \ref{fig1} as a function of $b$ and $d$.

We can see that the decay of the energy as a function of $b$ with $k$=1 (blue line) is slower for the H$_{2}$ and H$_{4}$ molecules. In the case of LiH and BeH$_2$, $b$-cutoff values of $b$=0.7 and $b$=0.4, respectively, are enought to obtain an error between points of the order of 10$^{-5}$, while for H$_2$ we should choose $b$=10 to obtain an error of the order of 10$^{-3}$. While we can decrease the error by increasing the $b$-cutoff, this will impact in the number of integration intervals needed as we will see in the next sections.

Increasing the iteration step to $k$=10 reflects in a faster decay as a function of $b$ (green line) in the H$_2$ and H$_4$ systems, which allows to choose a smaller $b$-cutoff. On the other hand, the $b$-cutoff for LiH and BeH$_2$ does not decrease at $k$=10 when compared to the $k$=1 results.

The H$_4$ case is particular because of the presence of a minimum corresponding to the ground state and a converged value at large $k$ that corresponds to an excited state. In this system, the $b$-cutoff is chosen to be the minimum value of $b$=9 for $k$=1 and $b$=6 for $k$=10. Increasing the iteration step to $k$=10 in this system produces a faster decay that shifts the minimum to a smaller $b$-cutoff.

Finally, note that the behaviour of the integration along $z$ for an iteration step of $k$=1 (orange line) is similar in all molecules and the $d$-cutoff can be set to $d$=4 in all the systems. In the case of LiH and BeH$_2$, however, increasing the iteration step to $k$=10 increases the value of the $d$-cutoff required for the integration along $z$. For example, to get the correct value at a $b$-cutoff of $b$=1 for LiH and BeH$_2$, it is necessary to increase the $d$-cutoff value to $d$=5 and $d$=6 (red lines), respectively, which will increase the minimum required number of polynomials for the integration along $z$ ($n_z$). Although for the current chosen iteration step of $k$=10, and for $b$-cutoffs of $b$=0.7 and $b$=0.4, we can retain a $d$-cutoff value of $d$=4 in both systems, there might be a minimum value of $k$ for which we will have to start increasing the $d$-cutoff. We will discuss this in a later section when we analyze the iteration power $k$. As a practical procedure for an unknown system, we think it's convenient to choose the initial $b$-cutoff based on the results for $k$=1, with a fixed $d$-cutoff=4.

\subsubsection{Legendre polynomial order}

We proceed to determine the order of the Legendre polynomials required to reach a convergence of the energy. In order to eliminate the error caused by the $z$ integration, we set the parameters to large enough values of $n_z$=100 and $k$=10 for all calculations in order to determine the cutoff for $n_y$, and we employed the Gauss-Legendre quadrature rule for the integration along $z$. We then compared the results as a function of $n_y$ by using both the trapezoidal method and the Gauss-Legendre quadrature rule. After choosing the cutoff for $n_y$, we determined the cutoff for $n_z$ by using both integration methods. The integration intervals $[0,b]$ and $[-d,d]$ were set according to the $b$- and $d$-cutoffs obtained in the previous section. For H$_2$, LiH, BeH$_2$ and H$_4$ systems, the $b$-cutoffs were set to $b$=10, $b$=0.7, $b$=0.4 and $b$=6, respectively, while the $d$-cutoff was $d$=4 in all cases.

\begin{figure*}[!t]
\center{\includegraphics[width=\columnwidth, keepaspectratio]
        {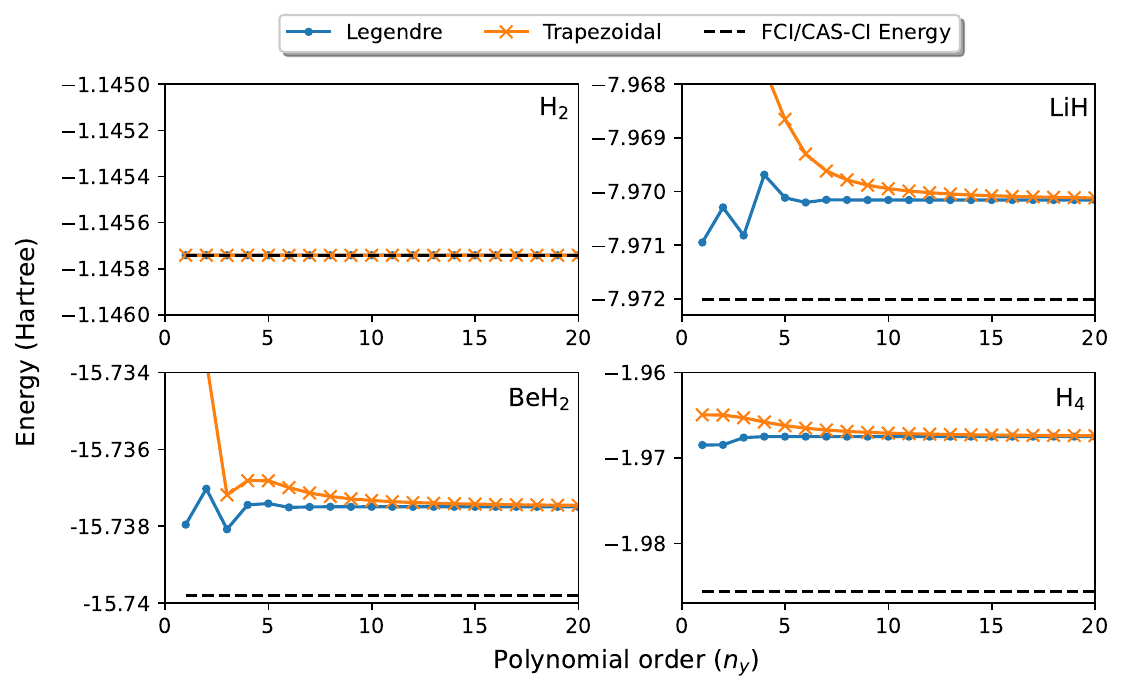}}
        \caption{Ground state energy as a function of the Legendre polynomials/discrete intervals for the integration along $y$ ($n_y$) at iteration step of $k$=10 for H$_2$ (top left), LiH (top right), BeH$_2$ (bottom left) and H$_4$ (bottom right). Integrations are performed using $b$-cutoffs of $b$=10, $b$=0.7, $b$=0.4, $b$=6, respectively, with the Gauss-Legendre quadrature rule (blue) and the Trapezoidal method (orange). The number of Legendre polynomials for the $z$ variable were set to n$_z$ = 100. The FCI energy for H$_2$ and H$_4$, and the CAS-CI energy for LiH and BeH$_2$, are represented as black dashed lines.}
        \label{fig2}
\end{figure*}

The energy results as a function of $n_y$ are shown in Fig. \ref{fig2} using Gauss-Legendre rule (blue line) and trapezoidal (orange line) integration methods. The true energy calculated using FCI and CAS-CI methods are shown in black dashed lines as a reference. We can see that for LiH, BeH$_2$ and H$_4$ we can choose a Legendre polynomial order of $n_y$=8, $n_y$=8 and $n_y$=4, respectively, with an error between points of the order of 10$^{-5}$. Although not shown with the current chosen figure scale, in order to obtain the same values using the trapezoidal integration method, the number of discrete intervals would have to be set to $n_y$=40, $n_y$=73 and $n_y$=190, respectively. Furthermore, when using Gauss-Legendre quadrature, we can observe that the energy values for $n_y$=1 are smaller than the converged value in all cases. This suggests that not only the convergence is faster when using Gauss-Legendre quadrature rule, but also that choosing a value to $n_y$=1 could provide a good approximation. The reason for this might be understood by looking at Fig. \ref{fig1} in the previous section. When choosing an appropriate $b$-cutoff, the faster the decay, the more we can approximate it as a polynomial of order 1 (linear decay). In the case of H$_2$, the system is too simple and the variation of the energy as a function of $n_y$ is of the order of 10$^{-9}$, which is negligible, but we choose $n_y$=15 as the converged value.

\begin{figure*}[!t]
\center{\includegraphics[width=\columnwidth, keepaspectratio]
        {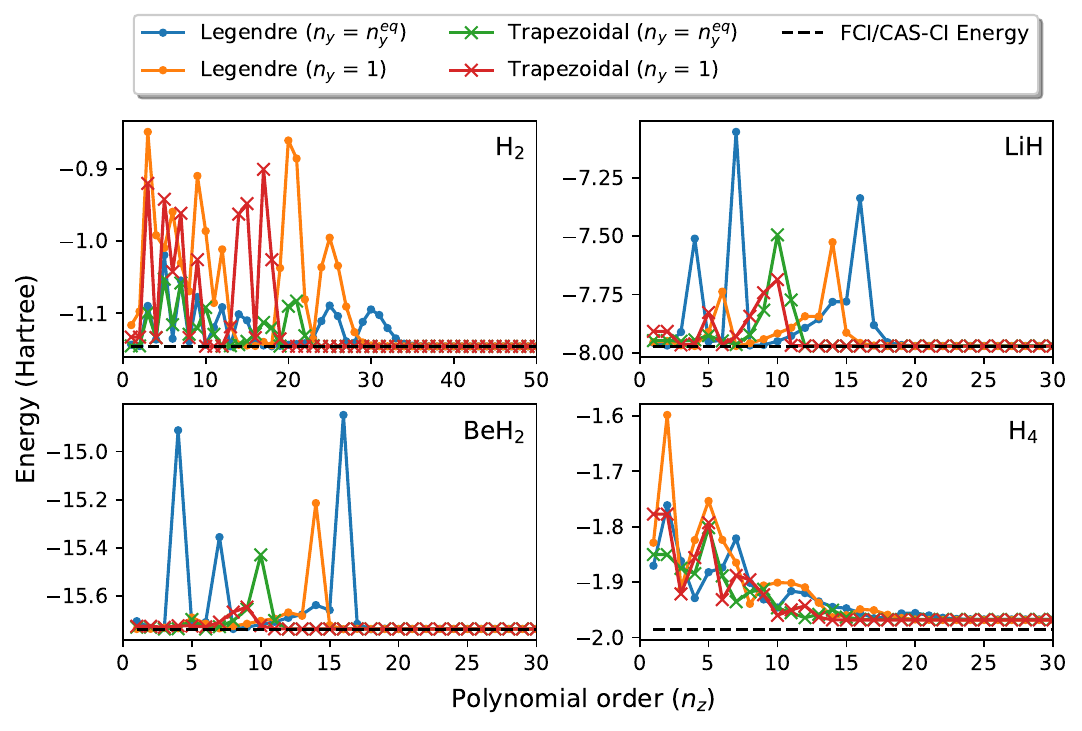}}
        \caption{Ground state energy as a function of the polynomial order/discrete intervals for the integration along $z$ ($n_z$) at iteration step of $k$=10 for H$_2$ (top left), LiH (top right), BeH$_2$ (bottom left) and H$_4$ (bottom right). Integrations are performed using the Gauss-Legendre quadrature rule with $n_y$=$n_y^{eq}$ (blue) and $n_y$=1 (orange), and the Trapezoidal rule with $n_y$=$n_y^{eq}$ (green) and $n_y$=1 (red). The $n_y$=$n_y^{eq}$ values are $n_y$=15, $n_y$=8, $n_y$=8, $n_y$=4, respectively, and correspond to the converged values chosen in Fig. \ref{fig2}. The FCI energy for H$_2$ and H$_4$, and the CAS-CI energy for LiH and BeH$_2$, are represented as black dashed lines.}
        \label{fig3}
\end{figure*}

We then analyze the variation of the energy as a function of $n_z$ for the chosen $n_y$ values using Gauss-Legendre quadrature rule (blue line) and trapezoidal integration (green line) methods in Fig. \ref{fig3}. We also compare these results to the ones obtained using a linear approximation with $n_y$=1 (orange and red lines, respectively).
We choose the convergence values of $n_z$ with an error tolerance of 10$^{-5}$ between two consecutive points. Once again, the true energies calculated using FCI and CAS-CI methods are shown in black dashed lines as a reference.

In the case of H$_2$, for $n_y$=15, we obtain values of $n_z$=38 and $n_z$=24 for Gauss-Legendre and trapezoidal integration, respectively. When choosing $n_y$=1, the values decrease to $n_z$=32 and $n_z$=21.
The LiH system, with $n_y$=8, results in values of $n_z$=22 and $n_z$=13, which decrease to $n_z$=20 and $n_z$=12, respectively, for $n_y$=1.
Next, the BeH$_2$ system values are $n_z$=22 and $n_z$=13 for $n_y$=8, decreasing to $n_z$=19 and $n_z$=12 for $n_y$=1.
Finally, for H$_4$ we get $n_z$=25 and $n_z$=17 in the case of $n_y$=4, which decreases to $n_z$=22 and $n_z$=15, respectively, for $n_y$=1.

We can observe that the general trend is that, for the integration along $z$, the trapezoidal method requires less discrete intervals than the number of Legendre polynomials required when using Gauss-Legendre quadrature. Furthermore, it seems that choosing $n_y$=1 not only results in a better approximation of the energy in Fig. \ref{fig2}, but also decreases the required number of $n_z$. The better performance of the trapezoidal method in this case is assumed to be due to the highly oscillating nature of the function $z$ after the integration in $y$, which is not well approximated by Legendre polynomials.

From these results, we can conclude that a hybrid Legendre-Trapezoidal method employing Gauss-Legendre quadrature along $y$ with n$_y$=1 and trapezoidal integration along $z$ seems to be the most convenient in order to reduce the number of operations required.

We want to point out that other types of quadrature rules\cite{gauß1814methodus} were explored as possible alternatives for the integration along $y$ and $z$. Gauss-Laguerre quadrature allows to treat the integration $\int_{0}^{\infty}dye^{-iyz\hat{H}}f(y)$ for the $y$ variable, while Gauss-Hermite approximates the integration $\int_{-\infty}^{\infty}dze^{-\frac{z^2}{2}}f(z)$ for the $z$ variable. In this case, the variables in Eq. \ref{gen_legendre} have to be refined as $y'_{j_y}=\frac{y_{j_y}}{1-y_{j_y}}$ and $z'_{j_z}=\frac{z_{j_z}}{1-z_{j_z}^2}$ in terms of the Laguerre polynomials $y_{j_y}$ and Hermite polynomials $z_{j_z}$ with the intervals $[0,\infty]$ and $[-\infty,\infty]$, respectively. The term $c_{a,b,c,d}=c_{y_{j_y}}c_{z_{j_z}}$ is now included in the summations where $c_{y_{j_y}}=\frac{1}{(1-y_{j_y})^2}$ and $c_{z_{j_z}}=\frac{1+z_{j_z}^2}{(1-z_{j_z}^2)^2}$ are the constants for the change of interval for $y$ and $z$, while $w_{j_y}$ and $w_{j_z}$ are the weights of the Laguerre and Hermite polynomials with order $n_z$ and $n_y$, respectively. The infinite integration intervals in these methods have the advantage of the suppressing the requirement of setting the cutoff intervals. However, infinite intervals reflects in the requirement of infinite polynomials for $n_y$ and $n_z$, making impossible to reach convergence.

\subsubsection{Iteration order}

\begin{figure*}[!t]
\center{\includegraphics[width=\columnwidth, keepaspectratio]
        {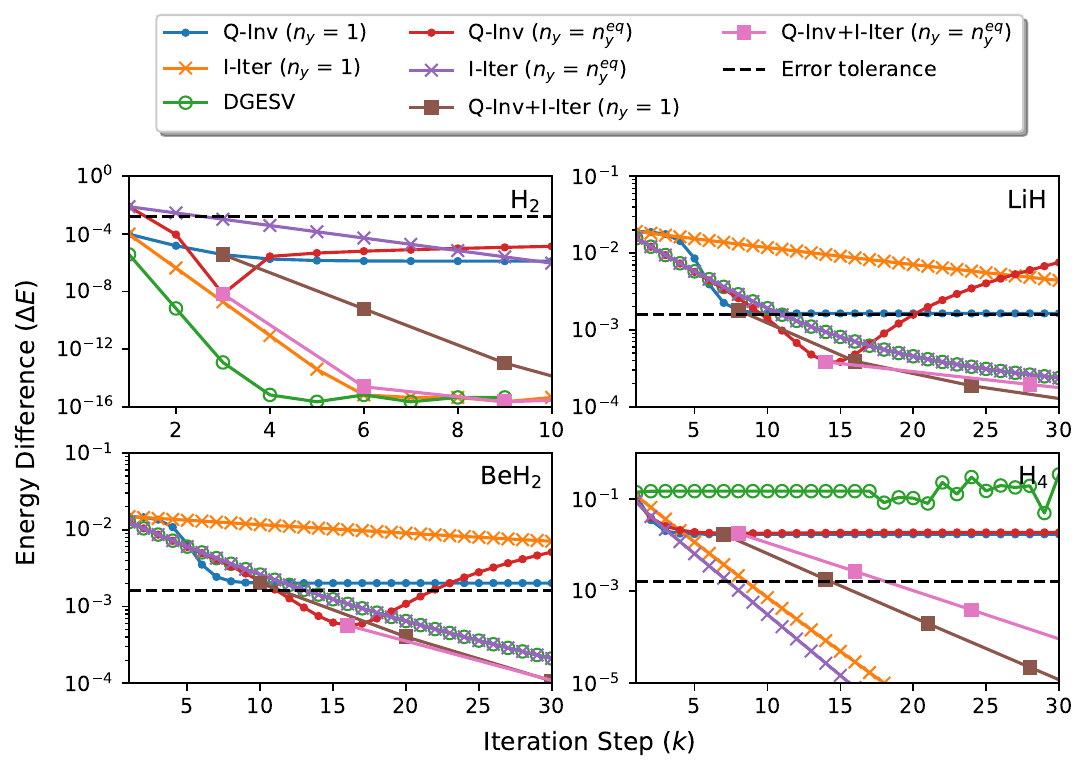}}
        \caption{Energy difference ($\Delta E$) between the estimated energy and the FCI/CAS-CI energy in logarithmic scale as a function of the iteration order $k$ for H$_2$ (top left), LiH (top right), BeH$_2$ (bottom left) and H$_4$ (bottom right). The blue and orange lines are the results for the quantum inverse algorithm (Q-Inv) and the inverse iteration method (I-Iter), respectively, with $n_y$=1. The red and purple lines are the result of the Q-Inv and I-Iter with the chosen cutoff for $n_y$ ($n_y^{eq}$), respectively. The brown and pink lines are the results employing I-Iter after employing Q-Inv up to a certain iteration step $k$ and denoted as Q-Inv+I-Iter. The green lines are the energies obtained by calculating the inverse using the DGESV method. Finally, the black dashed lines are error tolerance based on the chemical accuracy (1.6$\times$10$^{-3}$).} 
        \label{fig4}
\end{figure*}

After setting the initial set of parameters used for the integration, we can analyze the behaviour of the Q-Inv method by calculating the energy as a function of the iteration power $k$. 

We plot the error between the calculated energy (logarithmic scale) and the exact energy obtained using FCI/CAS-CI for each of the systems ($\Delta E$) in Fig. \ref{fig4}, where the chemical accuracy (1.6$\times$10$^{-3}$ Hartree) is set as the error tolerance (black dashed line). The quantum inverse method (Q-Inv) is performed for the $n_y$=1 (blue line) and $n_y$=$n_y^{eq}$ (red line) conditions, with their respective $n_z$ values. The $n_y^{eq}$ refers to the converged value chosen in the previous section for each system within a specified tolerance. 

We can observe that, when choosing $n_y$=1, the error decreases with increasing $k$ until it reaches a convergence value, while, when choosing $n_y$=$n_y^{eq}$, we notice the presence of a minimum value at different $k$ in all systems (not noticeable in H$_4$ due to the scale), which has a lower error compared to the $n_y$=1 case. The presence of a minimum is consistent with the reported results in Ref \citenum{kyriienko2020kyriienko} and, as the detailed analysis of the integration parameters showed in previous sections, it seems that this behaviour may be the consequence of the limitation of the $d$-cutoff parameter with increasing iteration step $k$. For a fixed $d$-cutoff, there is a maximum $k$ for which the parameter holds before introducing error. After a certain value of $k$, the $d$-cutoff may need to increase accordingly to maintain the accuracy of the method, followed by a recalculation of the other integration parameters as well, which is unpractical in the study of unknown systems where one does not posses the true energy value.

In the H$_2$ and LiH systems, the $n_y$=1 condition is enough to guarantee energy values below or equal to the error tolerance, respectively, but results in higher errors for the BeH$_2$ and H$_4$ cases. The minimum obtained in the $n_y$=$n_y^{eq}$ condition provide results with errors below the tolerance for H$_2$, LiH and BeH$_2$, but higher for H$_4$.

For comparison, we also calculate energy values employing the I-Iter method using the first order inverse ($\hat{H}^{-1}$) for the same $n_y$=1 (orange line) and $n_y$=$n_y^{eq}$ (purple line) conditions, and the iteration employing the exact inverse obtained by lower-upper (LU) decomposition (DGESV) value (green line). In this case, we see that, with exception of the H$_2$ system, the I-Iter method with the $n_y^{eq}$ condition provides a faster decay of the error with increasing $k$. This behaviour overlaps with the Q-Inv results for their respective conditions at small values of k. The Q-Inv method then shows a faster decay on the error compared to the I-Iter method up to the converged and minimum values for the $n_y$=1 and $n_y$=$n_y^{eq}$ cases, respectively. The DGESV value, while providing better results for H$_2$, it overlaps with the I-Iter results for the $n_y$=$n_y^{eq}$ condition in the LiH and BeH$_2$ systems. More importantly, in the H$_4$ case, the DGESV calculation, in which the integration cutoffs cannot be chosen, fails to reproduce the values of the ground state energy and converges to an excited state. For this particular system, by analyzing Fig. \ref{fig1}, we can see that the Q-Inv and I-Iter methods allow the possibility of obtaining energies corresponding to different states by selecting a different $b$-cutoff.

The Q-Inv method performs better than the I-Iter method in terms of computational cost by decreasing the number of iteration steps by $k$-times. Furthermore, it also provides better accuracy up to a certain iteration power $k$. However, in some systems such as H$_4$, the error of the energy might be higher than the allowed tolerance. In that case, we suggest the combination of the Q-Inv and the I-Iter methods. We can employ the Q-Inv method to obtain the minimum values and then apply the I-Iter method using that value of $k$ to further decrease the error while minimizing the iteration scaling. We denote this as Q-Inv+I-Iter and it is shown on each system for the $n_y$=1 (brown line) and $n_y$=$n_y^{eq}$ (pink line) conditions. The constant values for $n_y$=1 are obtained at $k$=3, $k$=8, $k$=10 and $k$=7 for H$_2$, LiH, BeH$_2$ and H$_4$, respectively, with an error between consecutive points of the order of 10$^{-5}$, while the minimum values for $n_y$=$n_y^{eq}$ are located at $k$=3, $k$=14, $k$=16, $k$=8. We observe that, after minimization, only 1 to 2 iteration cycles are required in order to obtain values below the error tolerance, and that the decay seems to be faster for the $n_y$=1 case, with exception of the H$_2$ system.

We note that the results on this work are obtained by assuming an initial Hartree-Fock state which might not be the best available option. By employing different methods to estimate an initial state closer to the ground state energy (increasing the initial wavefunction overlap) they performance of the method might be improved.\cite{fomichev2024arrazola, ge2018cirac,dong2022tong, lin2020tong}

\section{Conclusions}
In this work we analyzed the performance of the quantum inverse (Q-Inv) method for the H$_2$, LiH, BeH$_2$ and H$_4$ systems as a function of the integration parameters and the iteration power $k$. We performed the integrations employing the trapezoidal method and the Gauss-Legendre quadrature rule, and compared the results to the inverse iteration (I-Iter) method and the exact inverse (DGESV).

Results showed a constant $d$-cutoff in all systems when $k$=1 and a dependence of the $d$-cutoff with increasing iteration power $k$, which resulted in higher requirement of $n_z$. 
A combined Legendre-Trapezoidal integration method allowed the reduction in the number of matrix multiplications required. The Gauss-Legendre quadrature rule seemed to work better for the $y$ integration by reducing the number of Legendre polynomial $n_y$. Particularly, the $n_y$=1 case results provided a good approximation of the energy. On the other hand, the trapezoidal method seemed to work better in the $z$ integration by reducing the number of discrete intervals $n_z$.

The Q-Inv method provided better results than the I-Iter method up to a certain iteration power $k$, at which reaches a minimum value. The Q-Inv+I-Iter method can be employed by applying the I-Iter method after Q-Inv minimization to further decrease the error at reduced computational cost. 

According to these results, for a general system, we suggest a procedure that involves the calculation of the initial $b$-cutoff at $k$=1 for a fixed $d$-cutoff=4, followed by a Legendre-Trapezoidal integration method for the optimization of $n_z$, using Gauss-Legendre quadrature rule for $y$ with $n_y$=1 and trapezoidal integration for $z$. Then, proceed to find the minimum value of the Q-Inv method as a function of the iteration power and finally apply the I-Iter method using the obtained $k$ for an established tolerance.

\section*{Acknowledgements}
This work was supported by JSPS KAKENHI (JP23H01921),
JST-FOREST Program (JPMJFR221R), JST-CREST Program (JPMJCR23I6), and
MEXT Q-LEAP Program (JPMXS0120319794).

\bibliography{QComp.bib}
\end{document}